\documentclass[fleqn,usenatbib]{mnras}

\usepackage{newtxtext,newtxmath}

\usepackage[T1]{fontenc}

\DeclareRobustCommand{\VAN}[3]{#2}
\let\VANthebibliography\thebibliography
\def\thebibliography{\DeclareRobustCommand{\VAN}[3]{##3}\VANthebibliography}

\usepackage{graphicx}	\usepackage{amsmath}

\title[dSph oblateness from $\psi$DM perturbation]{Explaining the oblate morphology of dwarf spheroidals 
with Wave Dark Matter perturbations.}

\author[R. Della Monica et al.]{
Riccardo Della Monica,$^{1}$\thanks{E-mail: rdellamonica@usal.es}
Ivan de Martino,$^{1}$
Tom Broadhurst$^{2,3,4}$
\\
$^{1}$Universidad de Salamanca, Departamento de Fisica Fundamental, P. de la Merced, Salamanca, ES\\
$^{2}$Department of Physics, University of the Basque Country UPV/EHU, E-48080 Bilbao, Spain\\
$^{3}$DIPC, Basque Country UPV/EHU, E-48080 San Sebastian, Spain\\
$^{4}$Ikerbasque, Basque Foundation for Science, E-48011 Bilbao, Spain
}

\date{Accepted XXX. Received YYY; in original form ZZZ}

\pubyear{2023}

\begin{document}
\label{firstpage}
\pagerange{\pageref{firstpage}--\pageref{lastpage}}
\maketitle

\begin{abstract}
We investigate whether the oblate, spheroidal morphology of common dwarf spheroidal galaxies (dSph) may result from the slow relaxation of stellar orbits within a halo of Wave Dark Matter ($\psi$DM) when starting from an initial disk of stars. Stellar orbits randomly walk over a Hubble time, perturbed by the pervasive "granular" interference pattern of $\psi$DM, that fully modulates the dark matter density on the de Broglie scale. Our simulations quantify the level of stellar disk thickening over the Hubble time, showing that distribution of stars is predicted to become an oblate spheroid of increasing radius, that plausibly accounts for the morphology of dSph galaxies. We predict a low level of residual rotation remains after a Hubble time at the 1-3 km/s level, depending on orientation, that compares with recent claims of rotation for some well studied local dSph galaxies. This steady internal dynamical evolution may be witnessed directly with JWST for well-resolved dwarf galaxies, appearing more oblate with look back time and tending to small disks of young stars at high redshift.

\end{abstract}

\begin{keywords}
keyword1 -- keyword2 -- keyword3
\end{keywords}

\section{Introduction}

Stars formation in local dwarf galaxies is typically associated with rotating HI gas disks, whereas the classical dwarfs spheroidals (dSphs) are gas-free and have an oblate spheroidal morphology of old stars. This raises the question of how the majority of stars were formed in common dSph dwarfs if indeed stars are all born in disks, in particular for typical dSphs with simple single-age populations for which there is no sign of a significant merger evolution. A possible clue in this regard is the significant oblateness of the stellar distribution typically seen in the local dSph galaxies \citep{Carlsten2021}. Careful dynamical studies have often claimed dSphs have sizeable, kpc scale, dark matter (DM) dominated cores \citep{Walker2009, Moskowitz2020}, that contrast with the much more centrally peaked and stellar distributions of typical luminous elliptical galaxies of lower mass-to-light ratios indicating stars dominate the mass within the central few kpc and commonly show dynamical disturbances, including tidal tails and large stellar shells so that major merging explains their observed sphericity.

The Wave Dark Matter ($\psi$DM) interpretation of DM as an ultra-light boson ($m_a\sim10^{-22}$ eV) is physically motivated by the light axions of String Theory \citep{Arvanitaki2010, Hui2017, Luu2020} and arguably has attractive properties on galaxy scales resulting from the inherently large de Broglie wavelength ($\lambda\sim $ kpc) \citep{Hu2000, Schive2014} and provides an appealing solution for the cores as standing wave solitons of $\psi$DM in the ground state \citep{Schive2014}. Halos of pervasive interference are also predicted where the density is fully modulated, from zero to double the mean density, ranging from constructive to destructive interference \citep{Schive2014, Mocz2017, Veltmaat2018}. Lensing effects are now understood to be significant even in projection, significantly acting the relative magnification of images due to the interference pattern that causes the critical curves to become highly corrugated \citep{Chan2020}. The positions of lensed images near the critical curve can be perturbed  milli-arcseconds and with large flux anomalies expected of typically 30\%, with a log-normal tail to high magnification, that have been claimed to quantitatively match the flux anomalies of well-studied lensed QSO's, on galaxy scales \citep{Amruth2023}.
  
Here we explore a causative relation between the stellar distribution and the dominant DM in dSph galaxies in relation to the newly appreciated relaxation process predicted for $\psi$DM , where the  mass distribution of the halo is inherently granular leading to a constrained random walk for stellar orbits \citep{Schive2014, Hui2017}. Importantly, dynamical friction in this dSph context has been highlighted by \citet{Hui2017} in relation to the long recognised "timing problem" that several ancient globular clusters orbiting the Fornax dwarf at a surprisingly large radius of several kpc, whereas the dynamical friction of any form of heavy particle DM is firmly expected to have brought these GC's to the center of Fornax within the Hubble time \citep{Tremaine1975, Tremaine1976}. An important realization by \citet{Hui2017} is that for $\psi$DM there is no overdense "wake" behind a heavy prohibiting mass, as sufficiently light bosons cannot be concentrated on less than the de Broglie scale so that dynamical friction is far less than particle DM. 

The analytical work of \citet{Church2019} and \citet{BarOr2019} have explored the plausibility of accounting for the measured increase of stellar scale height and age of local stars in the disk of the Milky Way, showing that the scattering effect of $\psi$DM substructure is significant. At the center of the Milky Way there may also be evidence of a solitonic core on a scale of 100 pc \citep{DeMartino2020} from the enhanced velocity dispersion of bulge stars \citep{Portail2017} in this region and the rapid rotation of a thin disk of stars on this scale \citep{Schonrich2015}.

 Firstly, we outline our simulations of the halo of $\psi$DM, and our initial disk in Section \ref{sec:simulations} then we present results, showing the evolution of the vertical scale height and of the radial size and oblateness of the stellar population over time, and the resulting residual rotation that may persist over a Hubble time in Section \ref{sec:results}. Finally, our conclusions are presented in Section \ref{sec:conclusions}.

\section{Simulations of stellar scattering in dwarf galaxies}
\label{sec:simulations}

\begin{figure*}
    \includegraphics[width = \textwidth]{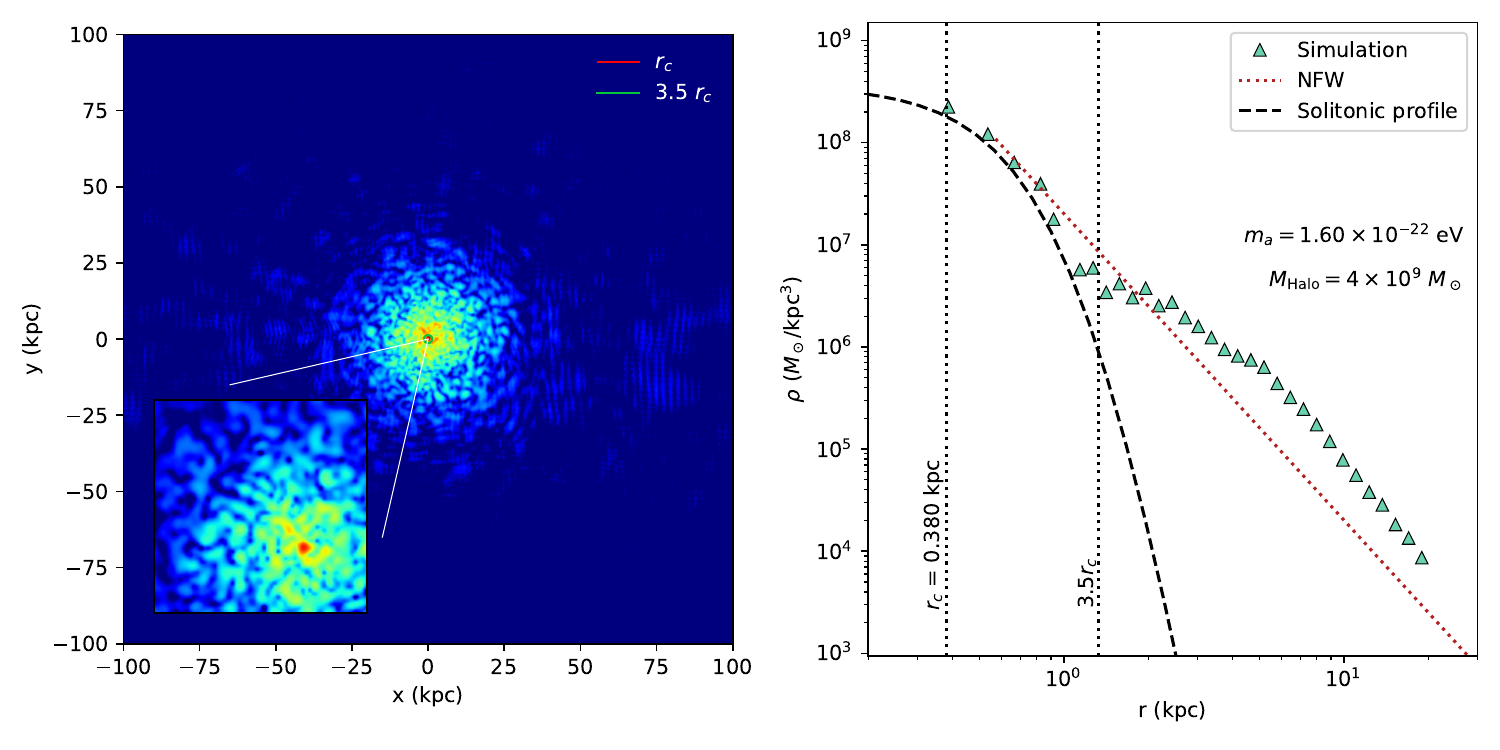}
    \caption{The density field that we obtained as a result of our merger simulation. \emph{Left panel:} we report a slice on the plane $z = L/2$ of the squared-norm of the wavefunction $\psi$. The colour map corresponds to the density values in the logarithmic scale. The red and green circles are centered on the central solitonic overdensity of the halo and their radii correspond to the core radius $r_c$ (red circle) and $3.5r_c$. The latter coincides with the points where the solitonic profile breaks and the halo is dominated by $\psi$DM fluctuations. \emph{Right panel:} we report the radial density profile of the halo in logarithmic scale (green triangles). The dashed black line reports the best-fitting soliton profile in Eq. \eqref{eq:soliton} in the inner regions of the halo. Vertical dotted lines report the best-fitting core radius $r_c$ and $3.5$ times this value. For greater radii, the density profile of our simulation departs from the solitonic profile and follows a NFW-like $\rho\propto r^{-3}$ (red dotted line).}
    \label{fig:sim_profie_density}
\end{figure*}

\subsection{The Wave Dark Matter halo}
The density distribution of ultra-light bosons ($m_a\sim10^{-22}$ eV) is well-described by the solution of the system of the non-relativistic Schrödinger-Poisson (SP) equations
\begin{align}
    i\hbar\frac{\partial}{\partial t}\psi(\vec{r}, t) &= -\frac{\hbar^2}{2m_a}\nabla^2\psi(\vec{r},t)+m_a\Phi(\vec{r},t)\psi(\vec{r},t),\label{eq:sp_1}\\
    \nabla^2\Phi(\vec{r}, t) &= 4\pi G|\psi(\vec{r},t)|^2,\label{eq:sp_2}
\end{align}
where $\Phi(\vec{r},t)$ is the gravitational potential and the wavefunction $\psi(\vec{r},t)$ is normalized so that its norm, $|\psi(\vec{r},t)|^2$, corresponds to the mass density of the $\psi$DM particles, $\rho(\vec{r},t)$. This comes under the assumption that, since all the particles in the halo share the same wavefunction, the probability density related to such wavefunction, $|\psi|^2$, traces the physical mass density of $\psi$DM \citep{Mocz2017}. The SP system of equations can be recast in a fluid representation \citep{Suarez2011,Chavanis2011} via the Madelung transformations allowing for an effective description of the $\psi$DM system as a fluid with density and velocity depending on the phase of the complex wavefunction. The dependence of the phase from the position can lead to the presence of turbulence generated by spatial wave interference that produces vortex lines in the field \citep{Mocz2017}. As describing this phenomenon analytically can be quite cumbersome, the best approach relies on the numerical integration of Eqs. \eqref{eq:sp_1} and \eqref{eq:sp_2} under appropriate boundary and initial conditions. 
The pioneering numerical simulations of the SP equations by \citet{Schive2014} have demonstrated the existence of a groundstate solution in the form of a solitonic core that is surrounded by wavelet fluctuations generated by the interference pattern of the $\psi$DM wavefunction. The density profile of such a solution is well approximated by the formula
\begin{align}
    &\rho_{\rm sol}(r) = \rho_0\left(1+9.1\times10^{-2}\left(\frac{r}{r_c}\right)^2\right)^{-8},\label{eq:soliton}
\end{align}
with the central density being
\begin{align}
    &\rho_0 = 3.1\times10^{6}\left(\frac{m_a}{2.5\times10^{-22}\textrm{ eV}}\right)^{-2}\left(\frac{r_c}{\textrm{kpc}}\right)^{-4}\frac{M_\odot}{\textrm{kpc}^3}\,,
\end{align}
where $r_c$, namely the core radius, is the radius at which the density drops by a $\sim$ half w.r.t. its peak value. The solitonic core portrays the ability to form a cored configuration thanks to an internal quantum pressure generated by the Heisenberg uncertainty principle over De Broglie wavelength scales which supports the system against gravitational collapse \citep{Hui2021}. This allows the density profile to flatten in the central region. The soliton solution in Eq. \eqref{eq:soliton} coincides with the spherically symmetric stationary solution to SP equations that is known as boson star \citep{Liebeling2012} and has been demonstrated \citep{Schive2014, Mocz2017} to be a rather general final state of $\psi$DM simulations, basically independent of the initial conditions, that constitutes the virialized endpoint of any $\psi$DM distribution after relaxation has occurred. In accordance with this result, we perform three-dimensional numerical simulations of $\psi$DM halos by letting evolve an initial state composed by a fixed number $n_c$ of stationary solitons with fixed radius $r_c$, randomly distributed in a box with side $L = 200$ kpc. We thus fix the initial state $\psi(\vec{r},0)$ of the system to be a superposition of different profiles as described by Eq. \eqref{eq:soliton}. The wavefunction is discretized on a cubic grid of dimension $N^3 = 512^3$ (thus ensuring that at least 4 points per linear dimension are enclosed within each soliton core radius \citep{Schwabe2016, Mocz2017}) and is evolved by means of a second order pseudo-spectral solver, which provides second-order convergence in time and spectral convergence in space. This method is based on a ``kick-drift-kick" scheme which allows to obtain the wavefunction at a time $t+\Delta t$ by applying a unitary operator to the wavefunction at time $t$. Namely,
\begin{equation}
    \psi(\vec{r},t+\Delta t) = e^{-\frac{i}{\hbar}H\Delta t}\psi(\vec{r},t)
    \label{eq:wf_evolution}
\end{equation}
where $H$ is the system's Hamiltonian. This can be split into the kinetic ($K$) and potential ($U$) terms in the following way:
\begin{equation}
    e^{-\frac{i}{\hbar}H\Delta t} = \underbrace{e^{-\frac{i}{\hbar}U \frac{\Delta t}{2}}}_{\textrm{kick}}\cdot\underbrace{e^{-\frac{i}{\hbar}K{\Delta t}}}_{\textrm{drift}}\cdot\underbrace{e^{-\frac{i}{\hbar}U \frac{\Delta t}{2}}}_{\textrm{kick}}.
\end{equation}
Basically, the wavefunction $\psi(\vec{r}, t)$ is first evolved by a half-step ($\Delta t/2$) kick by the potential operator. The result of this operation is then evolved by a full-step kinetic drift and, finally, a last half-step potential kick completes the evolution from $t$ to $t+\Delta t$. The kinetic and potential operators, on the other hand, are defined by means of projections back and forth the Fourier space for the wavefunction $\psi$. In particular, by denoting $\mathcal{F}$ the direct Fourier transform operator, by $\mathcal{F}^{-1}$ the inverse Fourier transform and by $\vec{k}$ the wave vectors corresponding to the space positions $\vec{r}$:
\begin{itemize}
    \item The kinetic operator $K = p^2/2m_a$ is expressed in Fourier space as $K = \hbar^2k^2/2m_a$ and it acts on the Fourier-transformed wavefunction, $\hat{\psi} = \mathcal{F}[\psi]$.
    \item The potential operator is given by $U(\vec{r}, t) = m_a\Phi(\vec{r}, t)$. Although this function is expressed in the real spatial domain, and hence acts on the wavefunction $\psi$ itself, it is derived by the Poisson equation in Eq. \eqref{eq:sp_2} which, on turn, is solved by a projection in the Fourier space. More specifically, the Laplacian operator is expressed as $\nabla^2\to -k^2$ and hence, the Fourier-transformed potential $\hat{\Phi} = \mathcal{F}[\Phi]$ can be easily determined by
    \begin{equation}
        \hat{\Phi}(\vec{k}, t) = -4\pi G\frac{\mathcal{F}[|\psi(\vec{r},t)|^2]}{k^2}.
    \end{equation}
    Finally, the potential in real-space domain is obtained with an inverse Fourier transform $\Phi(\vec{r}, t) = \mathcal{F}^{-1}[\hat{\Phi}(\vec{k}, t)]$.
\end{itemize}
Since the wavefunction is discretized on a regular grid, the operators $\mathcal{F}$ and $\mathcal{F}^{-1}$ are implemented into our algorithm in their discretized counterparts as 3D Fast Fourier Transform (FFT) and Inverse Fast Fourier Transforms (IFFT) \citep{FFT}. Moreover, as done in \citep{Schwabe2016}, we applied a Courant-Friedrichs-Lewy (CFL) condition on the timestep $\Delta t$ to guarantee the stability and accuracy of our simulations. This criterion is based on the requirement that the Hamiltonian operator, by which we evolve our wavefunction in Eq. \eqref{eq:wf_evolution}, changes the phase of $\psi$ by less than $2\pi$ and hence
\begin{equation}
    \Delta t\leq \max\left[\frac{m_a}{6\hbar}(\Delta x)^2, \frac{\hbar}{m_a|\Phi|_{\rm max}}\right]
\end{equation}
where $\Delta x \equiv L/N$ is the spatial resolution of our grid and $|\Phi|_{\rm max}$ is the maximum of the gravitational potential on the grid. Finally, let us remark that, since we assume periodic boundary conditions and since from the SP equations it follows the conservation of the total mass $M$ of the halo, differently from \citep{Schwabe2016,Du2018} the systems that we simulate are constantly perturbed (reheated) by the reflecting waves at the box boundary. This doesn't allow the system to reach perfectly the solitonic equilibrium configuration that is found analytically, but is closer to the cosmological scenario, in which waves coming from distant halos arrive from all directions, similarly to what is done in \citet{Schive2014} and \cite{Mocz2017}.
We fix the initial total mass $M$ of the halo by selecting a boson mass $m_a$, the number $n_c$ and the radii $\{r_{c,i}\}_{i=0}^{n_c}$ of the initial solitons that constitute the initial conditions of the system. In particular, the total mass carried by one soliton as described by the profile in Eq. \eqref{eq:soliton} is given by \citep{Schive2014}
\begin{equation}
    M_s(r_c) \approx 2.2\times 10^{10}{\displaystyle\left(\frac{m_a}{10^{-23}\textrm{ eV}}\right)^{-2}\left(\frac{r_c}{\textrm{kpc}}\right)^{-1}}M_\odot,
    \label{eq:total_mass}
\end{equation}
and hence the total (conserved) mass of our simulations is given by:
\begin{equation}
    M = \sum_{i= 0}^{n_c} M_s(r_{c,i}).
\end{equation}
Since our aim is to simulate halos that could be hosted by dwarf galaxies, we have fixed $M\sim 10^9 M_\odot$ and chosen different configurations of $n_c$ and $\{r_{c,i}\}_{i=0}^{n_c}$ accordingly, by verifying that the final state of the simulations is basically independent of the particular choice of the initial conditions, provided that the total masses are comparable. Moreover, the boson mass has been fixed to $m_a = 1.6\times10^{-22}$ eV. In Fig. \ref{fig:sim_profie_density} we report the resulting density field of our merger simulation, which was evolved for 10 Gyr. The radial density profile is characterized by a central solitonic core with $r_c = 380$ pc and an outer NFW-like ($\rho\propto r^{-3}$) profile, with the break occurring at $\sim 3.5r_c$. The density profile, as expected, reaches a cored configuration in the inner part of the halo, with a central density of $\rho_0 \sim 3 \times 10^{8}\,M_\odot/\textrm{kpc}^3$, which appears as the brighter over/density at the centre of the left plot, enclosed in the green ($r = 3.5r_c$) and red ($r= r_c$) circles. Outside this region, the halo is dominated by quantum fluctuations due to the interference pattern of the wavefunction.

\subsection{Heating of a stellar population}

The intrinsic granularity within galaxy halos related to the interference pattern of the $\psi$DM wave function on the de Broglie scale is predicted to perturb stellar orbits \citep{BarOr2019, Church2019}. We performed orbital dynamics simulations in order to quantify the level of disk thickening over the Hubble time. Our approach is based on the numerical integration of single stellar orbits in the gravitational potential well of the $\psi$DM halo obtained from our numerical simulation. We have considered a population of independently-evolving stars that do not interact with each other but only with the $\psi$DM halo. For each star in the population, consider its position, velocity, and acceleration at a given time $t_i$, given by $\vec{x}_i \equiv \vec{x}(t_i)$, $\vec{v}_i \equiv \vec{v}(t_i)$ and $\vec{a}_i \equiv \vec{a}(t_i)$, respectively. To evolve the star's position and velocity we have applied a first-order semi-implicit (energy-preserving) Euler method. More specifically, the updating rule at $t_{i+1} = t_{i}+\Delta t$ is given by
\begin{equation*}
    \left\{
        \begin{array}{l}
            \vec{v}_{i+1} = \vec{v}_{i}+\vec{a}_{i}\Delta t,\\
            \vec{x}_{i+1} = \vec{x}_{i}+\vec{v}_{i+1}\Delta t.\\
        \end{array}
    \right.
\end{equation*}

Accelerations are computed directly from the potential, which, in turn, is computed from the density field of the $\psi$DM halo. In particular, we have obtained the gravitational potential on the grid of our simulation by
\begin{equation}
    \Phi(\vec{x}, t) = -4\pi G\mathcal{F}^{-1}\left[\frac{\mathcal{F}[\rho(\vec{x}, t)]}{k^2}\right].
\end{equation}

Then, we have interpolated the function $\Phi(\vec{x}, t)$ by means of a regular grid linear interpolator in order to evaluate it at any position within the grid volume. Finally, the acceleration of the star at position $\vec{x}_i$ has been computed by $\vec{a}_i = -\nabla \Phi$. The precision of the integration depends on the integration time-step $\Delta t$ and on the space increment $\Delta x$ used for computing the gradient of the potential. Both these quantities have been set adaptively. More precisely, at each step of integration, we have computed the characteristic size of the $\psi$DM wavelets in that region $\lambda_{\rm char}$, and we have chosen $\Delta x$ to be a fraction of the wavelets $\Delta x = \lambda_{\rm char}/N $ (with $N\sim 10^2$) and the time-step of integration such that the distance travelled by the particle is a fraction of the wavelet size $|\vec{v}_i|\Delta t\sim \lambda_{\rm char}/M$ (with $M = 2$). This implies that at least two integration points for each wavelet fly-by of the star are taken into account (guaranteeing that the star does not surpass the wavelet without feeling its close interaction). For each star in the population, we have evolved its orbit for a time of $T = 10$ Gyr. 

Cooling gas during the formation era is predicted to be attracted by the large soliton over-density in the $\psi$DM halo centre and is therefore assumed to be cored in that region. Following this idea, stars have been initially placed on a thin disk following an exponentially decaying profile with a central Gaussian core. In particular, we have considered $N_{\rm stars} = 10^4$ randomly distributed over a disk of thickness $h = 100$ pc and whose number density follows the radial profile given by
\begin{equation}
    n(r) = n_0 e^{-\frac{r}{r_0}} *\mathcal{N}(r, r_c).
    \label{eq:stellar_radial_profile}
\end{equation}

Here $n_0$ is a normalization factor and the $*$ represent a convolution between an exponentially decaying profile with scale length $r_0$ (that we have fixed to be 500 pc so that the stellar density would drop by over 90\% on the typical scales of dwarf galaxies) and a gaussian normal distribution with amplitude coinciding with the soliton core radius $r_c$. This ensures smoothing in the inner regions of the exponential profile on the scales of the soliton radius.
Therefore, for each star, we have extracted a random value for the initial $z$ coordinate from a uniform distribution in the interval $[-h/2,\,h/2]$, a random value for the on-disk azimuthal angle $\varphi$ from the uniform distribution in the interval $[0,\,2\pi]$ and a random value for $r$ from the distribution in Eq. \eqref{eq:stellar_radial_profile} treated as a probability distribution. Finally, for each star, an initial velocity has been assigned with orientation on the equatorial plane in the tangential direction, and the value corresponding to the circular velocity at radius $r$,
\begin{equation}
    V(r)=\sqrt{\frac{G\mathcal{M}(r)}{r}},
\end{equation}
$\mathcal{M}(r)$ being the enclosed mass of the $\psi$DM halo at the star's initial position.

\section{Results}
\label{sec:results}

\begin{figure}
    \centering
    \includegraphics[width = \columnwidth]{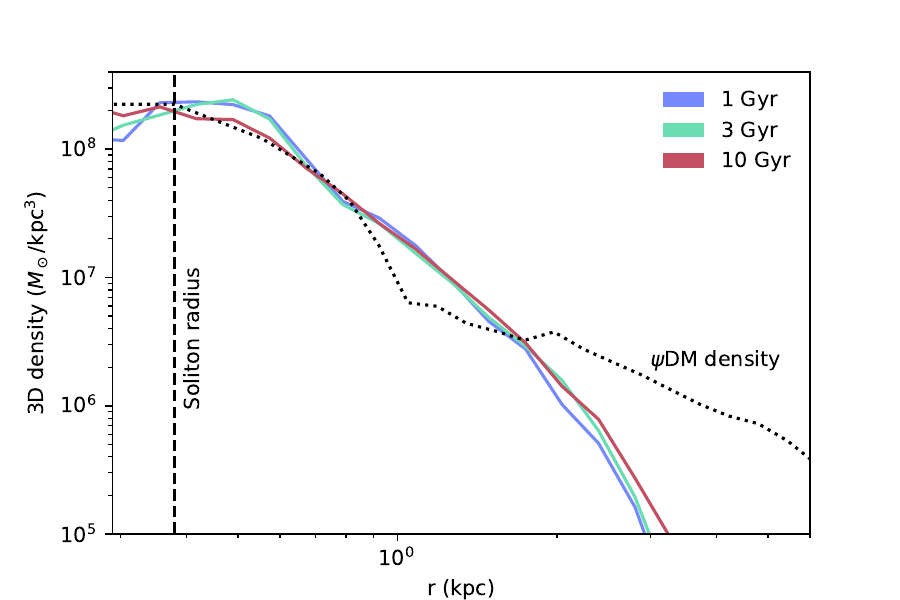}
    \caption{Mass density profiles of the stellar population after 1 (blue), 3 (green), and 10 Gyrs (red) of evolution in our simulation. The profiles have been rescaled to match the $\psi$DM density profile (dotted black line) in the central region. The dashed vertical line at 380 pc corresponds to the soliton radius in our simulation.}
    \label{fig:stellar_density}
\end{figure}

\begin{figure}
    \centering
    \includegraphics[width = \columnwidth]{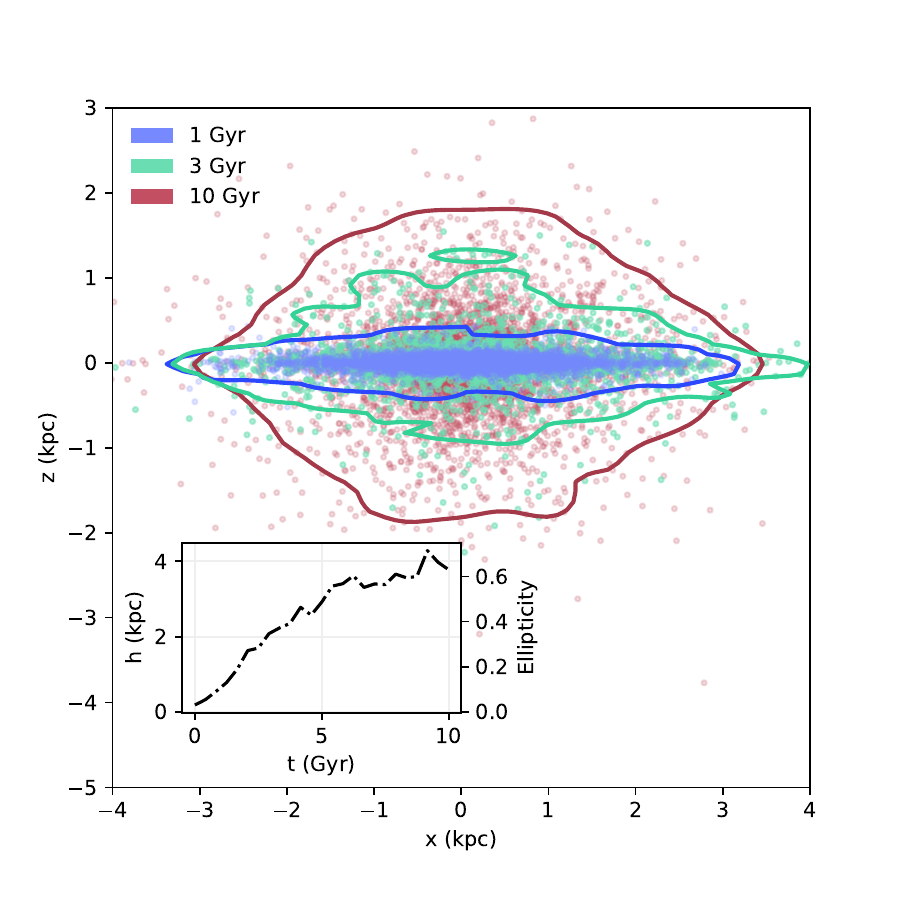}
    \caption{Stellar population positions on the $x-z$ plane after 1 (blue), 3 (green) and 10 Gyrs (red) of evolution. The dots correspond to the positions of the $10^4$ stars while the solid lines represent the contours enclosing 90\% of the stars at that epoch. The puffing-up of the stellar disk due to the orbital perturbation induced by the $\psi$DM granularity is clearly shown in our plot. The inset plot reports the growth over the Hubble time of the vertical spread of the stellar population and  of the ellipticity of the population.}
    \label{fig:disk_puff_up}
\end{figure}

\begin{figure}
    \centering
    \includegraphics[width = \columnwidth]{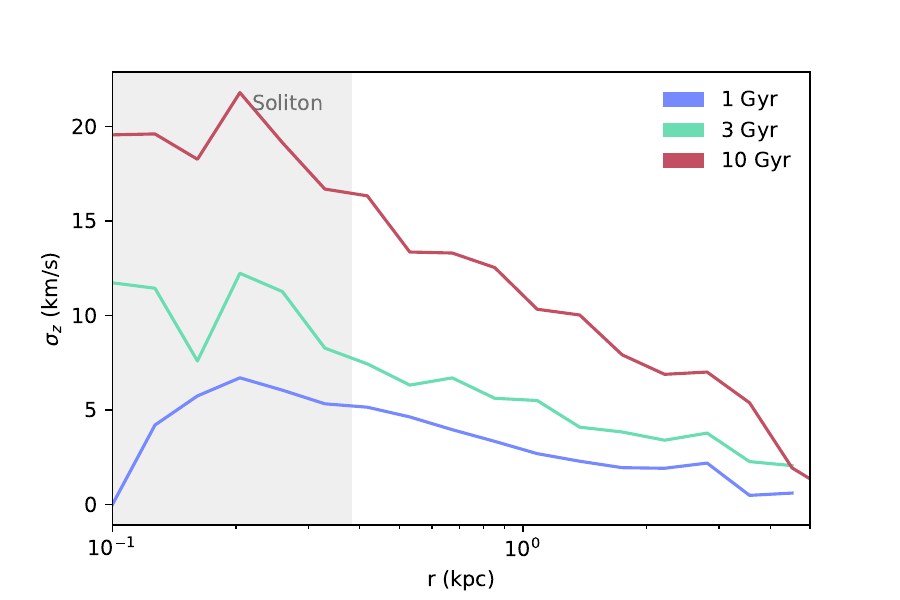}
    \caption{Vertical velocity dispersion profiles of the stellar population at 1 (blue), 3 (green) and 10 (red) Gyr, respectively. The shaded area corresponds to the region inside the soliton, where the peak of the profile. The population tends to acquire increasingly higher velocity dispersion perpendicularly to the disk as an effect of the interaction of the $\psi$DM granularity on the stellar orbits.}
    \label{fig:vertical_velocity_dispersion}
\end{figure}

In Fig. \ref{fig:stellar_density}, we report the evolution of the 3D stellar number density radial profile over time (after 1, 3 and 10 Gyrs of evolution). Only for the sake of visualization, the normalization factor for the stellar number density has been set so that stellar density at the soliton radius matches the $\psi$DM profile (depicted in the plot as a dotted line). The profiles exhibit very little to no evolution, implying that the stars do not tend to trace the $\psi$DM halo increasingly over time when looking at their number density. Their spatial distribution, on the other hand, shows evidence of a substantial evolution of the overall shape of the population in the vertical direction. Starting from an initial uniform height of 100 pc, at 1 Gyr reminiscence of the initial equatorial plane disk is still present, with the height of the population at $\sim600$ pc (computed as the symmetric interval containing 90\% of the stars in the vertical direction), corresponding to an ellipticity of 0.1 (computed as the ratio between the vertical and on-equatorial-plane horizontal spreads of the population). At 3 and 10 Gyr the stellar population is increasingly scattered by the interaction with $\psi$DM granules outside the equatorial plane, with the heights of the population reaching 2.2 kpc (0.37 ellipticity) and 3.8 kpc (0.63 ellipticity), respectively, approaching an oblate shape over the Hubble time (see Fig. \ref{fig:disk_puff_up}). 
This is reflected in the evolution of the vertical velocity dispersion of the stellar population. From an initial velocity field directed on the equatorial plane (implying a zero vertical velocity dispersion at the beginning of the simulation), the population tends to acquire increasingly higher velocity dispersion perpendicularly to the disk, with a profile that peaks at the soliton radius, as illustrated in Fig. \ref{fig:vertical_velocity_dispersion}.

\begin{figure}
    \centering
    \includegraphics[width = \columnwidth]{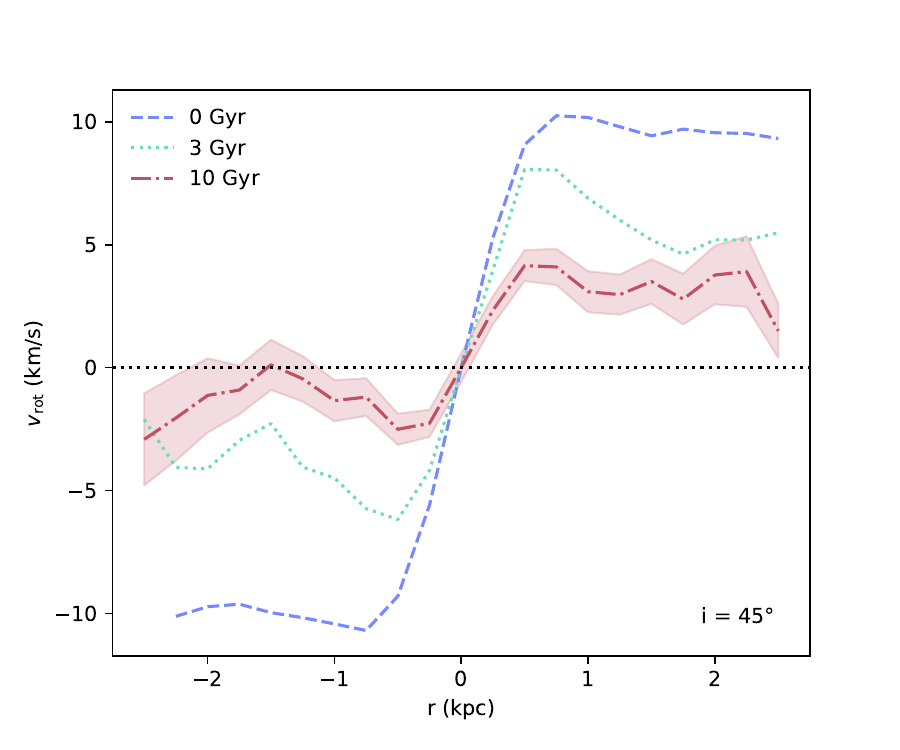}
    \caption{Rotation curve for an observer with inclination $i=45^\circ$ at the inital epoch (blue dashed line) and at 3 Gyr (green dotted line) and 10 Gyr (red dash-dotted line). The shaded area around the 10 Gyr profile corresponds to the statistical dispersion of the stellar velocity around the mean computed using a bootstrapping technique.}
    \label{fig:residual_rotation_curve}
\end{figure}

Finally, we have evaluated the possibility of detecting the residual rotation of the stellar disk at different epochs and for different angles of observation. In particular, we have considered a line of sight inclined by an angle $i$ with respect to the axis perpendicular to the initial disk and computed the rotation curve of the stellar population by considering the mean velocity of stars in radial bins (between $-2.5$ kpc and $2.5$ kpc with respect to the soliton) projected over this direction.
For the computation of the rotation curve at a given inclination angle $i$ we have fixed, without loss of generality due to the axial symmetry of the problem, an observer along the $yz$ plane at position $(0, \sin i, \cos i)$.  The line of sight of the observer, pointing towards the centre of the stellar population, is given by $\vec{k} = (0, -\cos i, -\sin i)$. We have thus split the stellar population in bins along the $x$ direction (\emph{i.e.} perpendicularly to the observer's line of sight) and for each star in the bin we have computed the line-of-sight velocity $v_{\rm los} = \vec{v}\cdot\vec{k}$. At this point, for each radial bin, we have a distribution of projected velocities on the line of sight $\{v_{\rm los, i}\}_{i = 0}^{N_{\rm bin}}$, where $N_{\rm bin}$ is the number of stars in the given bin. In order to minimize effects related to the reduced number of stars in each bin (especially towards the outskirts of the population), we have adopted a bootstrapping technique to compute the mean velocity and its statistical dispersion. In particular, in each bin, we have extracted $n = 100$ sub-samples each containing $k = N_{\rm bin}/3$ stars and we have computed the mean line-of-sight velocity $\bar{v}_{\rm los, j}$ for each sub-sample. This allowed us to compute a fixed-size velocity distribution for each sample $\{\bar{v}_{\rm los,j}\}_{j = 1}^n$. Performing the average over the obtained distribution we finally obtain  the rotation velocity at that given radial bin $v_{\rm rot} = \langle\bar{v}_{\rm los, j}\rangle$ and by performing the standard deviation of the distribution we obtain the statistical dispersion around the mean (which is reported as the shaded area in Fig. \ref{fig:residual_rotation_curve} for the 10 Gyr profile).

In Fig. \ref{fig:residual_rotation_curve} we report the resulting rotation curves for an inclination of $i = 45^\circ$ at the initial epoch and after 3 and 10 Gyr. While at the initial time, the stellar population displays a very well-defined rotation curve (with a flat profile at $\sim 10$ km/s for $|r|>1 $ kpc), at later epochs we can clearly see that the rotation curve reduces its amplitude in the outskirts of the galaxy, reaching 5 km/s at 3 Gyr and only $\sim 2$ km/s at 10 Gyr. This reduction is consistent with the picture that the $\psi$DM granularities perturb stellar orbits by converting the ordered rotational configuration at early epochs into a more chaotic out-of-the-plane motion, supported by an increase in the increase in the vertical velocity dispersion. Finally, in order to give even more support to our conclusions, we have computed the radial profile of the anisotropy parameter for the stellar population. This quantity is defined as \citep{Binney2008}
\begin{equation}
    \beta(r) = 1-\frac{\overline{v^2_t}(r)}{2\overline{v^2_r}(r)},
\end{equation}
where $\overline{v^2_t}$ is the mean quadratic tangential velocity and $\overline{v^2_r}$ the mean quadratic radial velocity computed in the same radial bins in which we have evaluated the rotation curve of the population. A value $\beta\to1$ means that the motion of the stellar population is dominated by a much larger spread in the radial direction than in the tangential one (radially biased). Conversely, a value $\beta\to-\infty$ implies a system that is dominated by ordered tangential motion (tangentially biased) and a value $\beta\to0$ corresponds to a system in which stellar motion is isotropic.
\begin{figure}
    \centering
    \includegraphics[width = \columnwidth]{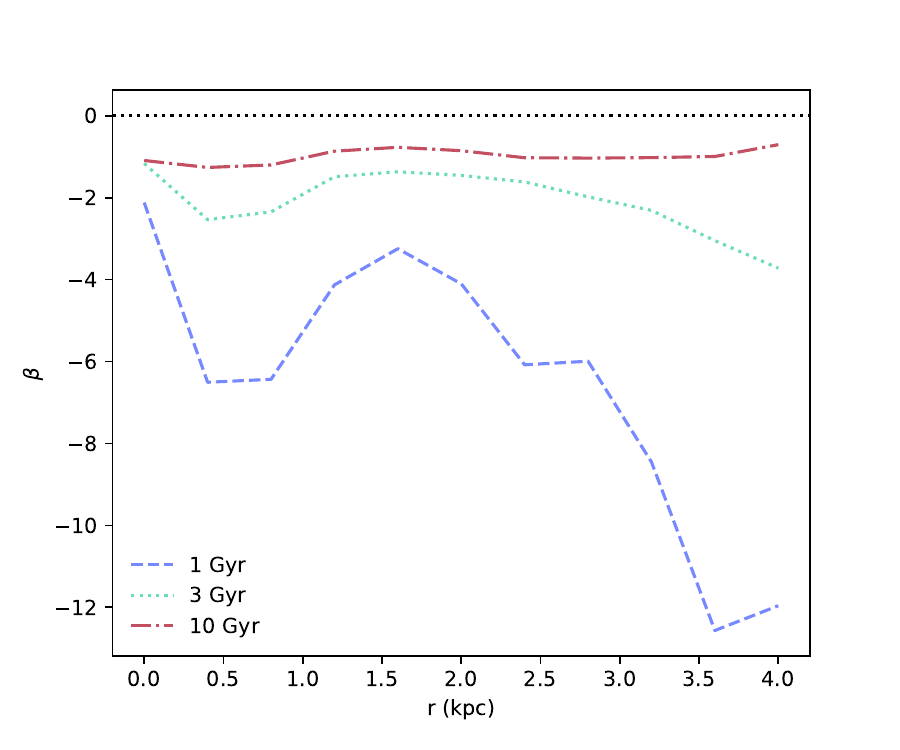}
    \caption{Radial profile of the velocity anisotropy parameter of the stellar population at 1 Gyr (blue dashed line), 3 Gyrs (green dotted line) and 10 Gyrs (red dash-dotted line). The figure shows that the system, whose values of the initial anisotropy parameter suggest a tangential bias (as expected by a rotating stellar disk), tends towards a more isotropic ($\beta \sim 0$) system (as expected by a pressure-supported kinematic of the stellar population). }
    \label{fig:anisotropy}
\end{figure}
In Fig. \ref{fig:anisotropy} we report the resulting anisotropy profiles at the three different epochs studied before. The results are in agreement with the general picture that emerges from our study: at earlier epochs the stellar motion is mostly an ordered rotation (accounting for the strongly negative $\beta$ profile at around -2 in the inner region and going down to -12 in the outskirts), reminiscent of the initial conditions on the galaxy disk, that is gradually superseded by disordered dispersion-supported behaviour at later epochs caused by the orbital perturbation induced by $\psi$DM granularities. This makes the anisotropy profile at 10 Gyr reach much closer values to 0, although not losing entirely the tangential bias, due to the residual rotation shown in Fig. \ref{fig:residual_rotation_curve}. Moreover, it is worth to notice, from the 3 Gyr profile, that the change in the anisotropy profile has a pretty clear radial dependence with the inner region tending towards $\beta\to 0$ at a faster rate than the outskirts.

\section{Discussion and Conclusions}
\label{sec:conclusions}

We have examined here the plausibility that the oblate, spheroidal distribution of stars that characterises dSph galaxies 
 may be accounted for by the level of orbital scattering predicted for $\psi$DM, due to the pervasive interference of 
DM substructure in this wavelike DM context. We have shown that a steady evolution from disk to spheroidal appearance may be
expected for our choice of initial conditions, that include an exponential stellar disk of kpc scale and for $\psi$DM 
with the standard favoured boson mass of approximately $10^{-22}$eV. The total mass of stars is  assumed to be negligible in comparison with the DM, to match the known high $M/L$ of the common dSph class of dwarf galaxy with masses of $10^{10}M_\odot$ and initial $v_{\rm max}$ chosen of $\simeq 30$km/s. We find that a smooth evolution is predicted to occur over a Hubble time for such DM dominated dwarf galaxies, evolving slowly to an oblateness that saturates over a Hubble time at nearly spheroidal level near $\sim 0.8$, with a modest residual rotation at the km/s level. This is encouragingly close to the new claims of rotation seen in some dSph galaxies \citep{MartinezGarcia2023}. For those dSphs that do not show a net rotation and where the stellar spectral data are of sufficient quality and quantity we would interpret this absence as a projection effect where the line of sign is more aligned with the rotation axis. Having recognised this prediction, a more through statistical analysis is now warranted. The qualitative behaviour is not very dependent on the precise choices of the main assumptions provided the 
galaxy is DM dominated and that the de Broglie wavelength is astronomical in scale. We also see a clear tendency towards modestly larger stellar scale radius over time, in qualitative agreement with the recent calculation of \citet{Chowdhury2023}, though an initial spherical Plummer profile was adopted in contrast to our assumption of an initial stellar disk.
Direct evidence for a flattened distribution
of stars can be seen in the well-resolved "UMi" dSph galaxy (see Figure 5 of \cite{Pace2020}), where a relatively young metal-rich population is clearly oblate, similar to our prediction at $\simeq 1$ Gyr (see Fig. \ref{fig:disk_puff_up}), compared with the more spheroidal distribution of older, metal-poor stars in UMi.

A clear implication at high redshift of the claim we have made here, in the context of $\psi$DM, is that the progenitor appearance of common dSph galaxies should appear more disk-like at high $z$. In principle, in this context, any young galaxy should look disk-like, rather than spheroidal within the first Gyr. At sufficiently high redshift, at $z>8$ all galaxies  should be less than 1 Gyr in age and the $\psi$DM simulations show they will not have suffered major mergers until much later or if at all given the suppression of low mass galaxies in this context \citep{Schive2016}. Hence, we predict that JWST should find that early dwarf galaxies should not appear to look spheroidal at early times but more disk-like with a possible solitonic core visible depending on gas cooling within the core region. The role of the soliton in modifying and enhancing our predictions in now warranted and may add to the halo scattering process that we have explored. The early formation of a soliton core is seen in the simulations formed early during the virialization process \citep{Schive2014} and hence may induce an initially relatively dense concentration of young stars within the soliton core.  Deep, well-resolved galaxies are being discovered at early times with JWST and indeed hints of central light concentration with older stars formed in an "inside out" manner with a distinct core have been discovered \citep{Baker2023} and also lensed examples with spectroscopic confirmation at $z>9$ have been discovered with well resolved elongated appearance, that than a smooth spheroidal form, encouraging further clarifying deep observations at high magnification \citep{RobertsBorsani2023, Williams2023}.

\section*{Acknowledgements}

RDM acknowledges support from Consejeria de Educación  de la Junta de Castilla y León. IDM acknowledges support from Grant IJCI2018-036198-I  funded by MCIN/AEI/10.13039/501100011033 and, as appropriate, by “ESF Investing in your future” or by “European Union NextGenerationEU/PRTR”. IDM and RDM also acknowledge support from the  grant PID2021-122938NB-I00 funded by MCIN/AEI/10.13039/501100011033 and by “ERDF A way of making Europe”. Finally, IDM acknowledges support from grant SA096P20 funded by Junta de Castilla y León.

\section*{Data Availability}

No new data were generated or analysed in support of this research.

\bibliographystyle{mnras}
\bibliography{biblio} 

\bsp	
\label{lastpage}
\end{document}